\documentclass[iop]{emulateapj}

\usepackage{natbib}
\usepackage{times}
\usepackage{color}
\usepackage{graphicx} 
\usepackage{hyperref}

\definecolor{RoyalBlue}{rgb}{0.25,0.41,0.88}
\definecolor{Red}{rgb}{0.61,0.12, 0.14}

\newcommand{\twCO}{\mbox{$^{12}{\rm CO}$}}       
\newcommand{\thCO}{\mbox{$^{13}{\rm CO}$}}       
\newcommand{\CeiO}{\mbox{${\rm C}^{18}{\rm O}$}} 
\newcommand{\kms}{\mbox{${\rm km\,s}^{-1}$}}

\shorttitle{CPD Observability}
\shortauthors{Perez et al.}

\begin{document}

\title{Planet formation signposts: observability of circumplanetary
  disks via gas kinematics}

\author{Sebastian Perez\altaffilmark{1,2}, A. Dunhill\altaffilmark{3},
  S. Casassus\altaffilmark{1,2}, P. Roman\altaffilmark{2,4},
  J. Szul\'agyi\altaffilmark{5}, C. Flores\altaffilmark{1,2},
  S. Marino\altaffilmark{1,2}, M. Montesinos\altaffilmark{1,2} }

\affil{$^1$ Departamento de Astronom\'ia, Universidad de Chile,
  Casilla 36-D, Santiago, Chile} 

\affil{$^2$ Millennium Nucleus ``Protoplanetary Disks'', Chile}

\affil{$^3$ Instituto de Astrof\'isica, Pontificia Universidad
  Cat\'olica de Chile, Vicu\~na Mackenna 4860, 7820436 Macul,
  Santiago, Chile}

\affil{$^4$ Center of Mathematical Modelling, Universidad de Chile.}

\affil{$^5$University of Nice-Sophia Antipolis, CNRS, Observatoire de
  la C\^ote d'Azur, Laboratoire Lagrange, F-06304, Nice, France}

\email{sperez@das.uchile.cl}

\begin{abstract}

The identification of on-going planet formation requires the finest
angular resolutions and deepest sensitivities in observations inspired
by state-of-the-art numerical simulations. Hydrodynamic simulations of
planet-disk interactions predict the formation of circumplanetary
disks (CPDs) around accreting planetary cores. These CPDs have eluded
unequivocal detection --their identification requires predictions in
CPD tracers. In this work, we aim to assess the observability of
embedded CPDs with ALMA as features imprinted in the gas
kinematics. We use 3D Smooth Particle Hydrodynamic (SPH) simulations
of CPDs around 1 and 5~$M_{\rm Jup}$ planets at large stellocentric
radii, in locally isothermal and adiabatic disks. The simulations are
then connected with 3D radiative transfer for predictions in CO
isotopologues. Observability is assessed by corrupting with realistic
long baseline phase noise extracted from the recent HL~Tau ALMA
data. We find that the presence of a CPD produces distinct signposts:
1) compact emission separated in velocity from the overall
circumstellar disk's Keplerian pattern, 2) a strong impact on the
velocity pattern when the Doppler shifted line emission sweeps across
the CPD location, and 3) a local increase in the velocity
dispersion. We test our predictions with a simulation tailored for
HD~100546 --which has a reported protoplanet candidate. We find that
the CPDs are detectable in all 3 signposts with ALMA~Cycle~3
capabilities for both 1 and 5~$M_{\rm Jup}$ protoplanets, when
embedded in an isothermal disk.
\end{abstract}


\section{Introduction}

Planets are expected to form during the evolution of circumstellar
disks of gas and dust \citep[e.g.,][]{Arm2011}. As a protoplanetary
core grows to approximately a Saturn mass, it becomes massive enough
to open a gap in the disk \citep{Lin1986, Lub1999}. This process is
the result of competition between gravitational, viscous and pressure
torques exerted onto the disk by the planet and by the disk itself
\citep{Cri2006}. The gap splits the disk into two radially distinct
zones. There are multiple detections of dust-depleted gaps and
cavities in well studied protoplanetary disks \citep{And2011,
  Seba2015}, but interestingly, no unambiguous detection of a forming
planet has yet been confirmed.

Hydrodynamical models of planet-disk interactions show that a single
giant protoplanet continues accreting from the outer disk at
formidable rates \citep[$\sim$2$\times$10$^{-4}$ $M_{\rm
    Jup}$yr$^{-1}$, ][]{Gre2013, Sha2013,Szu2014} even after its gap
is evacuated \citep{Kle1999,Pap2005}. Accretion streams converge onto
the vicinity of the giant developing a circumplanetary disk (CPD)
through which angular momentum disperses thus regulating planetary
accretion \citep{Lub1999, Ayl2009}.

In two-dimensional (2D) simulations, strong shocks appear near the
planet's Hill sphere leading to excessive inflow redirected towards
the planet, rapidly depleting the CPD material \citep{Lub1999,
  Dan2002}. These shocks appear much weaker in three-dimensional (3D)
calculations, leading to more persistent CPD structures
\citep{Bate2003, Ayl2009}. Grid-based simulations show that the inflow
of gas onto the protoplanet mostly happens in the vertical direction,
allowing material to cross the shock front near the Hill radius
\citep{Mac2010, Szu2014}. Similar results were found for MHD
simulations by \citet{Gre2013}. 


Most models show that CPD radii truncate to about 1/3 \citep{Ayl2009,
  Sha2013} or even 1/2 \citep{Gre2013, Szu2014} of the planet's Hill
radius. This implies that a Jupiter mass planet on a 100~AU orbit
could bear a CPD with a diameter of 4.5--7~AU. If located at 100~pc
away from Earth, the projected diameter translates into
$\sim$45-70~mas, within range of modern astronomical instrumentation.

Gas-giant protoplanet candidates have been detected embedded in the
HD~100546 and HD~169142 disks, two in each system \citep{Qua2013,
  Reg2014, Cur2014, Bil2014}. Both systems are young and nearby Herbig
Ae/Be stars ($<$10~Myr old, $<$145~pc) bearing large circumstellar
disks with confirmed gaps \citep{Oso2014, Wal2014}. The two directly
imaged detections are $L'$ (3.8$\mu$m) emission blobs: one is at 0.5''
angular separation ($\sim$52~AU at 100~pc) from HD~100546 \citep[also
  detected in $M'$ emission at 4.8$\mu$m,][]{Qua2014}, and the other
is at 0.16'' ($\sim$23~AU at 145~pc) from HD~169142. The compact
source in HD~169142 falls within a symmetric gap imaged in polarized
scattered light by \citet{Qua2013b}. Interestingly, there are no
near-IR counterparts for either candidate, supporting the idea that
these are in fact accreting gas giants with SEDs driven by
circumplanetary accretion \citep{Zhu2015}.

These CPDs have eluded unambiguous detection, mainly because of the
lack of predictions on CPD tracers. \citet{Ise2014} presented deep
continuum observations to detect a CPD around a protoplanet candidate
in LkCa\,15, with no positive results. CPD continuum emission may in
fact be scant since large dust grains ($>$100$\mu$m), probed by sub-mm
continuum observations, are sieved out of the planet-induced gap by
the outer disk pressure bump. Interestingly, only small grains
($<$100$\mu$m) enter the dust-depleted cavity \citep{Zhu2012} and make
it into the CPD. This dust-filtration is evidenced by the numerous
dust-depleted cavities seen in sub-mm observations
\citep{And2011}. CPD observational signposts to unambiguously confirm
forming planet candidates are scant. Realistic 3D simulations coupled
with radiative transfer are needed to predict the observational
signatures of circumplanetary disks at various wavelengths.

In this work, we connect 3D Smooth Particle Hydrodynamics (SPH) of CPD
dynamics (Section~\ref{sec:sims}) with 3D radiative transfer
(Section~\ref{sec:rt}) in the context of interferometric observations
of line emission. We aim to study the observability of CPDs as
features imprinted in the gas kinematics
(Section~\ref{sec:results}). We compute predictions tailored for the
Atacama Large Millimeter Array (ALMA) for common gas tracers
(Section~\ref{sec:hd100}). Implications and conclusions are discussed
in Section~\ref{sec:discussion}.

\section{Numerical Simulations} \label{sec:sims}


\begin{deluxetable}{lr}
  \tabletypesize{\scriptsize}
  \tablecaption{Simulations parameters}
  \tablewidth{0pt}
  \tablehead{
    \colhead{Parameter} & \colhead{Value}} 
    \startdata
  $R_{\rm in}$  & 0.35 $R_p$ \\
  $R_{\rm out}$  & 1.85 $R_p$ \\
  Initial $\Sigma$ profile  &  $\Sigma \propto R^{-1}$(*) \\
  Initial $H/R$ profile  &  0.05 (constant with $R$) \\
  Circumstellar disk masses$^\dagger$ &  \\
  ~~~Isothermal (SPH1, SPH2) & $5 \times 10^{-4} M_\odot$\\
  ~~~Adiabatic (SPH3)   &  $7 \times 10^{-5} M_\odot$
  \enddata
  \tablecomments{(*) normalized so that $R_p = 1$~AU would give $\Sigma = 100\,{\rm g~cm}^{-2}$ at $R = R_p$. ($\dagger$) for the nominal disk model presented in Sec.~\ref{sec:results}.}
\label{tab:sims}
\end{deluxetable}

We carried out a set of 3D SPH simulations to characterise the CPD
morphology and kinematics, which we feed into a radiative transfer
code (see Section~\ref{sec:rt}). We address the question of
observability signposts of a single snap CPD embedded in its parent
disk, after the simulations have reached a reasonably steady
state. Exploring how planet-disk interactions evolve with time is
beyond the scope of this paper.

\subsection{Three-dimensional SPH Simulations} \label{sec:sph}

\begin{figure*}
  \centering\includegraphics[width=\textwidth]{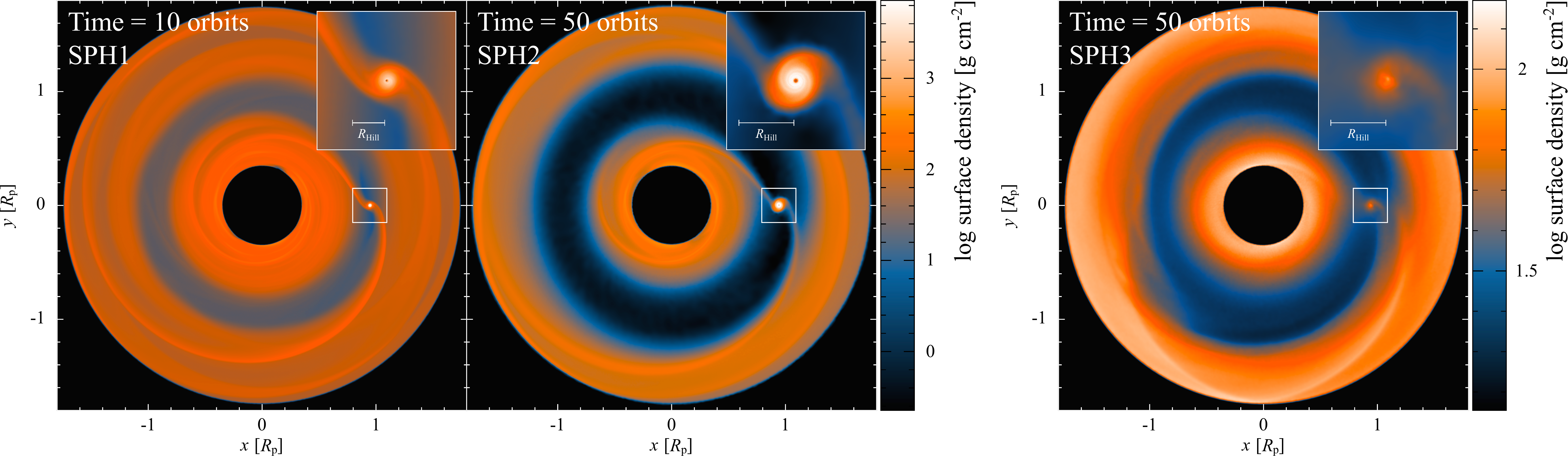} 
  \caption{Surface density maps after 10 and 50 orbits for the
    isothermal and adiabatic simulations listed in
    Table~\ref{tab:sims}. Axes are in planet orbital radius $R_{\rm
      p}$ units. {\it Left and middle} show isothermal runs for 1 and
    5~$M_{\rm Jup}$ planet candidates. {\it Right.} Adiabatic
    simulations for a 5~$M_{\rm Jup}$ (50 orbits). Upper right insets
    show a close up to the CPD kinematics. The planet bears a disk and
    drives a spiral wave without disrupting the disk structure
    heavily. Figures made using the SPLASH code
    \citep{Pri2007}. \label{fig:sph}}
\end{figure*}

We used a modified version of the SPH code {\sc Gadget-2}
\citep{Spr2005} to perform a suite of 3D simulations of a planet
embedded in a protoplanetary disk. The code has been modified to make
it more suitable for simulating planets embedded in disks
\citep[see][]{Dun2013}. We have simulated the full disk azimuth but a
restricted radial range.

We implement radial boundary conditions in a manner similar to
\citet{Ayl2010}, reducing the range of the simulated disk while still
providing high resolution around the planet. Although spiral waves
launched by the planet reflect off the boundary, they do not affect
the disk dynamics at the radius of the planet. A fuller description of
this method will be given in a forthcoming paper (Dunhill et al., in
prep). We summarise used disk parameters in Table~1.

To reduce the effect of transient waves on the planet at the start of
the simulation, we include an initial gap in the disk using the
prescription of \citep{Lub2006}. We model the planet as a point mass
potential with a sink radius, inside which any gas particles are
swallowed and their mass and momentum added to the planet. The
potential is unsoftened outside the sink radius. Initially, $R_{\rm
  sink} = 0.05\,R_p$ and decays exponentially to $R_{\rm sink} =
0.001\,R_p$ after approximately 8 orbits. This corresponds to $\sim$2
Jupiter radii for $R_p = 1$~AU.

We performed three different runs, two locally isothermal (SPH1 and
SPH2, where $T(R)$ is enforced so that $H/R$ remains constant) and one
with an adiabatic equation of state (SPH3, using an adiabatic index
$\gamma = 5/3$). It has been shown before that using an isothermal EOS
yields slightly larger circumplanetary disks than using more realistic
radiation physics \citep{Ayl2009}. SPH1 has a planet mass $M_p =
1~M_J$, while SPH2 and SPH3 have $M_p = 5~M_J$, all orbiting a
1~$M_\odot$ star.

For the simulations with $M_p = 5~M_{\rm Jup}$, we model the disk
using 2 million SPH particles. In order to achieve an equivalent
resolution within the CPD, we used 16 million SPH particles in
SPH1. We vertically resolve the CPD into $\sim 4$ SPH smoothing
lengths $h$ (typically $h \sim 10^{-3} R_p$ in the CPD midplane)
ensuring that we do not under-estimate the midplane density
\citep{Nel2006}. At this resolution, the artificial viscosity in the
simulations gives an effective Shakura-Sunyaev alpha parameter of
$\alpha \sim 0.005$ within the CPDs. Due to the extreme computational
expense, we halted SPH1 after 10 orbits of the circumstellar disk,
although lower-resolution tests indicate that it has settled into a
steady state by this time. SPH2 and SPH3 were halted after 50 orbits
(see Fig. 1).

These simulations are limited in that we neglect complex radiation
physics, including only viscous and shock heating but not passive
heating. However, the CPD structures we focus on are still present in
radiation hydrodynamic runs as shown by \citet{Ayl2009}.


\section{Radiative transfer predictions} \label{sec:rt}

The main driver of this investigation is to study under which
conditions an accreting protoplanet would be detectable through ALMA
observations of line emission. We have chosen bright and commonly
observed CO transitions which lie within the sub-millimeter (sub-mm)
range in ALMA. Rotational transitions of \twCO\ and the isotopologues,
\thCO\ and \CeiO, are known to trace the gas in protoplanetary disk
gaps and cavities \citep{Bru2013, Seba2015}. Most importantly, these
lines contain important kinematic information, essential to detect
companion objects embedded in the gas inside dust-depleted cavities.

We consider two model disks for our analysis. A nominal disk located
at 100~pc, inclined by 20$^\circ$ and hosting a planet at 100~AU, used
to illustrate the observational features revealed through CO
kinematics. The second model is tailored for the HD~100546 disk, with
an inclination of 42$^\circ$ and a 5~$M_{\rm Jup}$ planet candidate at
$R_{\rm p}=52$~AU \citep{Qua2014}. Our simulations are resampled
accordingly in Cartesian coordinates using a linear interpolation
scheme via SPLASH \citep{Pri2007}. The Cartesian cells are perfectly
cubic, with a cell size of 0.014$~R_p$, which, after scaling becomes
1.4~AU and 0.7~AU for $R_p=100~$AU and 52~AU, respectively.

For the isothermal runs (SPH1 and SPH2), temperature is an imposed
function of orbital radius, while for the adiabatic run (SPH3) we use
the SPH internal energies to calculate temperature by assuming a
standard mean molecular weight and adiabatic index. We scale the discs
using a $T(R) \propto R^{-1/2}$, consistent with measurements of
flaring discs \citep{Ken1987, And2011}. After scaling, temperatures at
100 AU reach $\sim$60 K for the isothermal disks and $\sim$2000 K for
SPH3, well above CO freeze-out (20-25 K).


In the event of CO freeze-out, a similar RT calculation can be applied
to species with enhanced abundances where CO is depleted, such as
DCO$^+$ or N$_2$H$^+$, or species that are formed by surface reactions
with CO ice, such as H$_2$CO, which have recently been detected in
disks at or beyond the CO snow-line \citep[][]{2013ApJ...765...34Q,
  2013Sci...341..630Q, 2013A&A...557A.132M}.

We compute synthetic images in CO(2-1) with the radiative transfer
code {\sc radmc3D} (Dullemond et al. 2015). Line radiative transfer is
done in LTE, using molecular data from the LAMDA
database\footnote{\url{http://www.strw.leidenuniv.nl/~moldata/}}. Line
widths include thermal broadening and a local (spatially unresolved)
microscopic turbulence set to a constant value of 0.1~\kms. We used a
fiducial molecular abundance relative of H$_2$ relative to \twCO\ of
10$^{-4}$. We adopted an ISM isotopic abundance $^{12}{\rm
  C}/^{13}{\rm C}$ of 76 \citep{Sta2008} and 500 \citep{Wil1994} for
\thCO\ and \CeiO, respectively.

Channel maps are rendered using {\sc radmc3d} ray-tracing. The results
are synthetic data cubes centered on the star, with a total width in
velocity of $\sim$16~\kms, and individual channels of 0.1~\kms. These
data cubes represent our sky model which is subsequently Fourier
transformed and resampled to ALMA's visibility plane.

For completeness, we calculate the continuum assuming a simple dust
distribution consisting of 30 per-cent amorphous carbon grains
\citep{Li1997} and 70 per-cent astronomical silicates \citep{Dra1984},
following the gas density. Grain size distribution follow a power-law
with exponent −3.5 from 0.05 to 1000~$\mu$m. We compute dust opacities
with Mie theory. CPD dust continuum predictions are addressed
elsewhere in the literature \citep[see, ][]{Wolf2005, Ise2014}, but
are strongly affected by gas pressure bumps \citep{Dan2015}. Continuum
emission in sub-mm observations arises from dust thermal radiation and
does not contain kinematic information, thereby, it is not within the
scope of this paper.

\begin{figure*}
  \centering\includegraphics[width=1\textwidth]{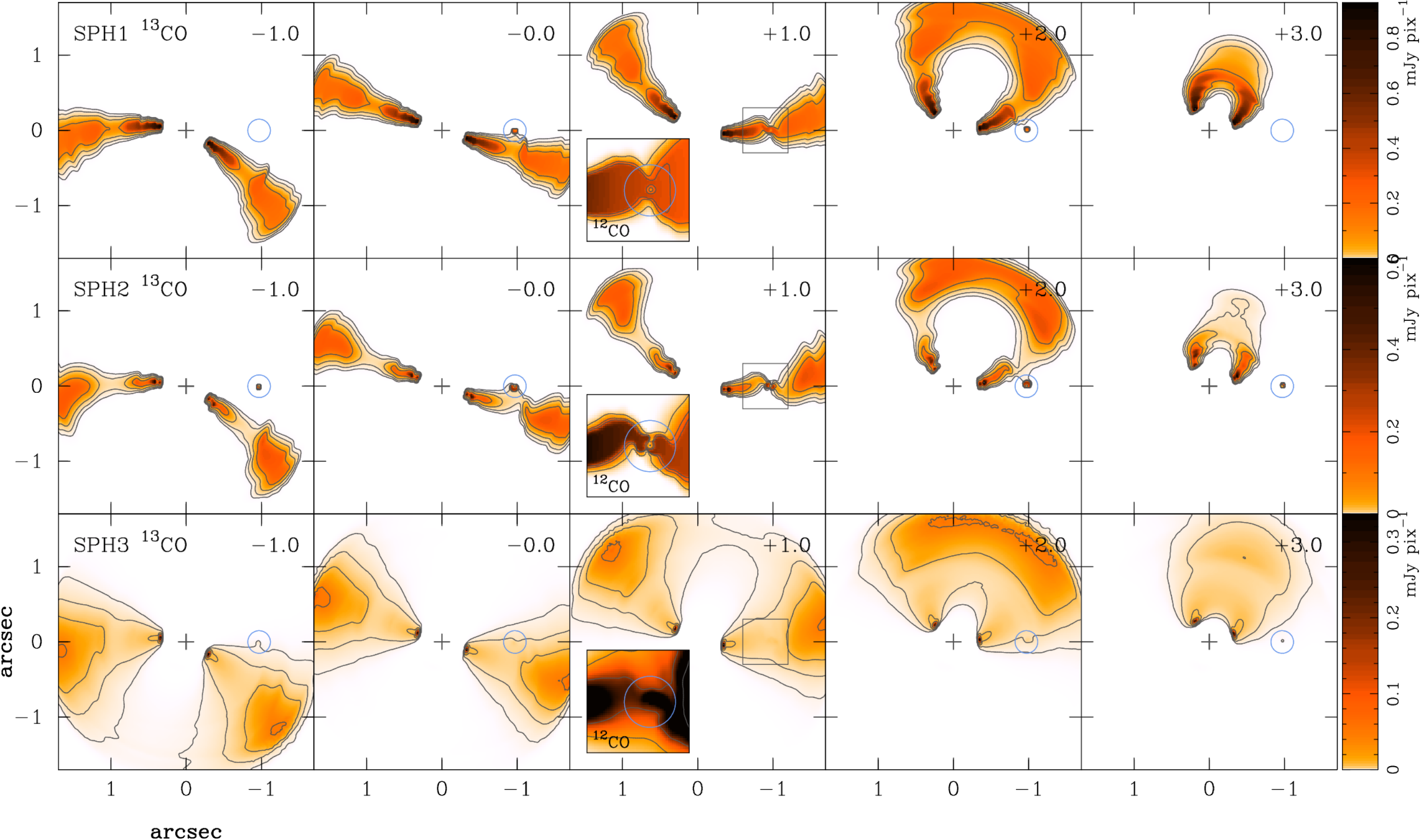}
  \caption{Predictions for \thCO\ emission based on SPH1, SPH2 and
    SPH3 (top, middle and bottom, respectively). The crosses show the
    central star and circles indicate the CPD position. From left to
    right, the maps velocities correspond to -1.0, 0.0, +1.0, +2.0 and
    +3.0~\kms. Channel widths are 0.1~\kms. Maps at
    0.0~\kms\ represent systemic velocity. Inset frames show
    predictions for \twCO. Color scale and contours are
    logarithmic. See Section~\ref{sec:results} for a full
    description. Fluxes are given in Jy~pixel$^{-1}$, where each
    synthetic pixel is 12~mas.  \label{fig:radtransfer} }
\end{figure*}

\begin{figure}
  \centering\includegraphics[width=0.75\columnwidth]{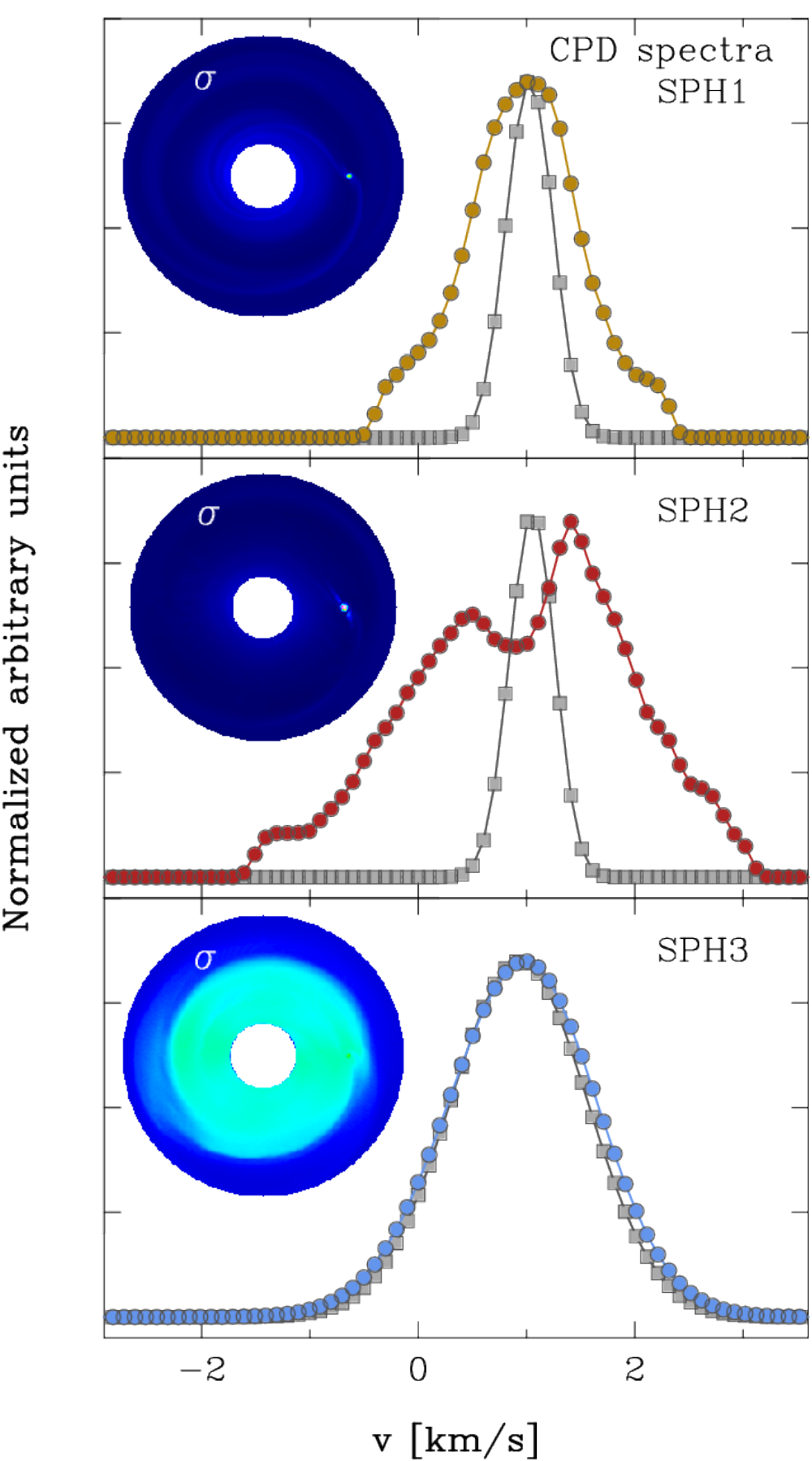}
  \caption{ \thCO\ CPD spectra extracted from an 80~mas aperture (in
    radius) centered on the planet (circle in
    Fig.~\ref{fig:radtransfer}). Top, middle and bottom spectra
    correspond to SPH1, SPH2 and SPH3, respectively. Squared data
    points (grey curve) are spectra extracted from the point symmetric
    location at the opposite side of the disk. Insets show second
    moment maps (velocity dispersion, $\sigma$) calculated over the
    region of interest.} \label{fig:spec}
\end{figure}

\section{Results and discussion} \label{sec:results}

Fig.~\ref{fig:radtransfer} presents our \thCO\ channel maps
predictions, at velocities -1.0, 0.0, +1.0, +2.0 and +3.0~\kms, for
the nominal disk model. Top and middle panels are isothermal
simulations for 1 and 5~$M_{\rm Jup}$ planets (SPH1 and SPH2,
respectively), while bottom panels show adiabatic results (SPH3).

We find that the presence of a CPD produces deviations from
circumstellar Keplerian kinematics. The circumplanetary disk is in
itself a miniature Keplerian accretion disk embedded in the gap, and
it can be separated in velocity from the overall Keplerian pattern of
the circumstellar disk. This velocity separation can be seen in
Fig.~\ref{fig:radtransfer} as compact emission at the CPD
position. This compact emission is persistent over a velocity range
given by the CPD kinematics.

Spectra extracted from the vicinity of the CPDs produces a broader
profile when compared with a spectrum extracted from the point
symmetric opposite side of the disk (see Fig.~\ref{fig:spec}). The
profile extracted from the vicinity of the 1~$M_{\rm Jup}$ planet
shows a broad core on top of even broader wings (top spectrum in
Fig.~\ref{fig:spec}). The middle panel in Fig.~\ref{fig:spec} shows
that the isothermal CPD around the 5~$M_{\rm Jup}$ planet exhibits a
distinct double-peaked profile, with a peak separation of
$\sim$1.5~\kms. Full CPD spectra span over $\Delta v
\sim$2.5~\kms\ for the isothermal 1~$M_{\rm Jup}$ CPD, while the
5~$M_{\rm Jup}$ covers more than $\Delta v \sim$4~\kms. The
double-peaked profile for the 1~$M_{\rm Jup}$ case is unresolved at
0.1~\kms\ resolution. The width of these spectral features may inform
on the size of their respective CPDs, and ultimately on the mass of
the accreting planet via their Hill radii. Fig.~\ref{fig:spec} also
shows that the CPD line wings for SPH1 and SPH2 end in an abrupt
shoulder. This is likely due to our planet accretion model; a point
mass with a sink radius inside which particles are swallowed and their
kinematics cannot be sampled.

The adiabatic CPD spectrum does not reveal distinctive features
(bottom spectra in Fig.~\ref{fig:spec}). The compact CPD emission is
also much less distinctive in the adiabatic disk (bottom panels in
Fig.~\ref{fig:radtransfer}). In opposition to the isothermal disk
where all compressive work is radiated away immediately, the adiabatic
disk cannot cool, reaching temperatues of $\sim$2000~K around the
CPD. This causes the hot gas to rapidly fill back the gap preventing a
CPD to fully develop (see second moment map in Fig.~\ref{fig:spec}),
which hinders clear identification of spectral features in an
adiabatic flow. Indeed, adiabatic and isothermal disks represent two
extrema of the phenomenon that we are modelling.

A second signpost of planet formation arises when the Doppler shifted
line emission of the circumstellar disk's Keplerian pattern sweeps
across the CPD location. The butterfly pattern becomes strongly bent
and twisted, while the point-symmetric location at the opposite side
of the disk remains undisturbed. This can be seen in the central
panels of Fig.~\ref{fig:radtransfer}. The insets show the same twisted
pattern but for \twCO. Optically thicker than \thCO, the \twCO\ maps
still reveal the kinematic bend, even for a 1~$M_{\rm Jup}$ planet
whose gap is shallower and its \twCO\ appear optically thick around
the CPD. For SPH3 the CPD vicinity is much hotter than in the
isothermal cases, producing enhanced \twCO\ emission (see bottom inset
in Fig.~\ref{fig:radtransfer}).

As noted in Section~3, our choice of temperature profile in SPH1 and
SPH2 is inconsistent with the scaling used for the RT
calculations. Self-consistency should result in a CPD approximately
twice as thick as in the SPH, reducing the midplane densities and
emitted fluxes by a similar factor. It is possible that the signatures
highlighted in Figures~2 and 3 would thus be at lower contrast to the
background, but still present at detectable levels as the velocities
are largely unaffected. CO freeze out at the CPD's location would be
prevented by adding a background temperature to account for accretion
radiation feedback \citep{Mon2015} and incident radiation from their
environment \citep[see][]{Sha2013}. Including self-consistent thermal
physics in future simulations will settle this discrepancy.

\subsection{HD~100546 through a 15 km baseline sub-mm observation}
\label{sec:hd100}

To assess the observability of these kinematic CPD signposts, we
performed a second calculation tailored for the protoplanet candidate
in HD~100546, based on the SPH2 run. We tied the fluxes of our model
to match approximately previous CO observations of this source
\citep{Wal2014, Pin2014}. We filtered our sky model using the
$uv$-coverage from the long baseline ($\sim$15~km) Science
Verification observations of HL~Tau \citep{alma2015}. We corrupted our
model with thermal and phase noises extracted directly from the HL~Tau
dataset. The simulated observation was then self-calibrated and
CLEANed using routines in the {\sc CASA} package.

\begin{figure}
  \centering\includegraphics[width=\columnwidth]{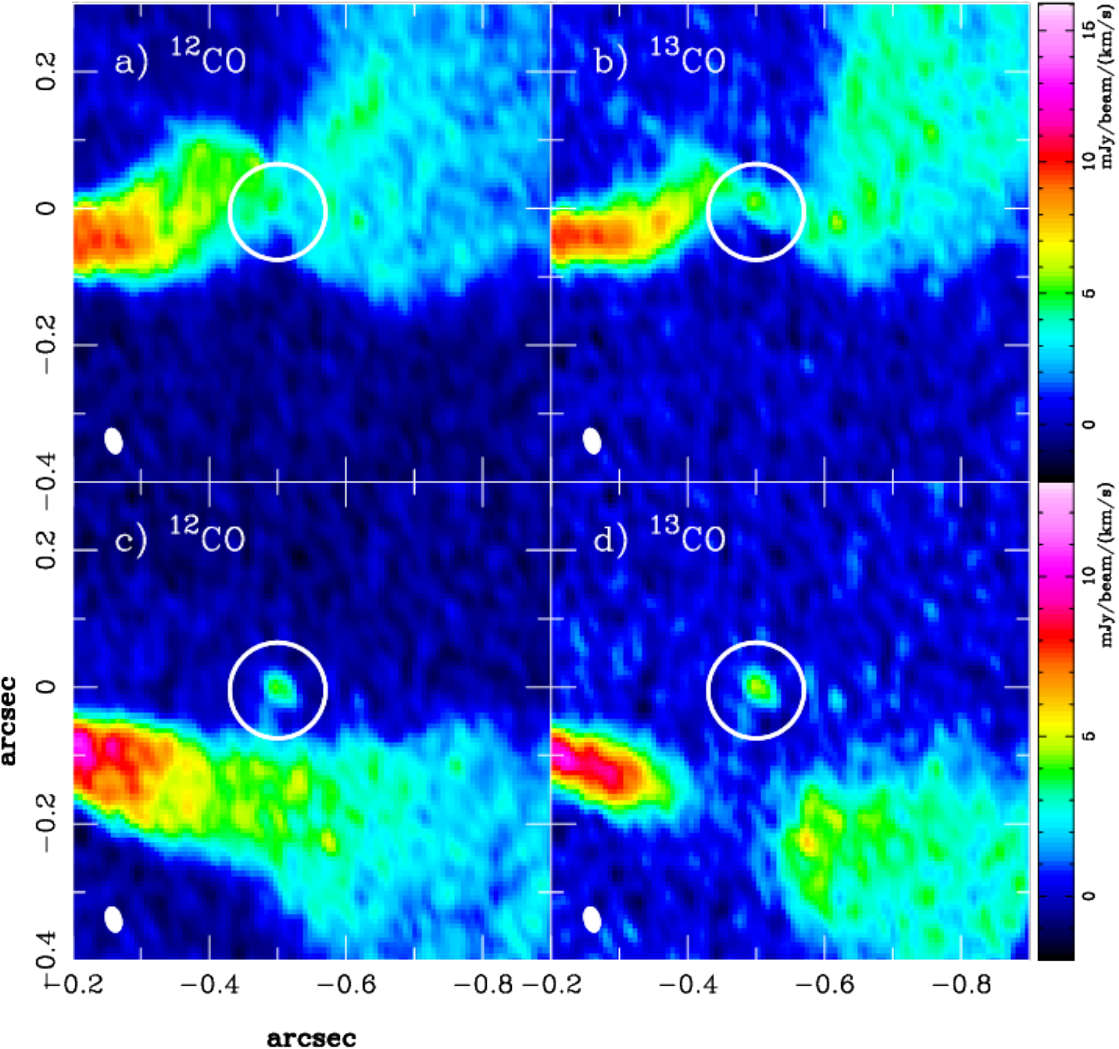}
  \caption{Channel map signpost predictions for HD~100546's CPD
    candidate. Calculation based on the 5~$M_{\rm Jup}$ isothermal
    SPH2 run for two channels centered at systemic velocity (bottom
    panels) and +2~km/s (top panels). Continuum subtracted $^{12}$CO
    and $^{13}$CO(2-1) emission after corrupting by HL~Tau's phase and
    thermal noise. a) and b) show the twisted Keplerian butterfly
    pattern, while c) and d) illustrate the striking CPD emission
    offset from the global Keplerian pattern. Channel widths are
    binned to 0.5~\kms. The rms noise is $\sim$1~mJy~beam$^{-1}$ in
    each bin}.  \label{fig:prediction}
\end{figure}

Fig.~\ref{fig:prediction} shows the HD~100546 simulated ALMA
observation. Left and right panels show selected channels for
\twCO\ and \thCO\ emission. The upper panels illustrate the recovery
of the bent Keplerian locus, while lower panels show how the CPD
compact emission can also be detected. Both emission features are
recovered at the 5~$\sigma$ level (rms noise is
$\sim$1~mJy~beam$^{-1}$).

\section{Conclusions} \label{sec:discussion}

The presence of a CPD produces distinct signposts in simulated CO
maps: a striking compact emission separated in velocity from the
overall Keplerian pattern of the circumstellar disk; a strong
influence on the velocity pattern of the gas when the Doppler shifted
line emission sweeps across the CPD location; and a local increase in
the velocity dispersion. Moreover, for the locally isothermal
simulation with a 5~$M_{\rm Jup}$, the CPD spectra even exhibits a
double-peaked profile. These distinctive features rely on kinematics
and can reveal the presence of an embedded CPD perturber even in
optically thick tracers like \twCO.

The feasibility of an ALMA observation of HD~100546 was assessed by
corrupting our synthetic visibilities with realistic phase and thermal
noises extracted from the HL~Tau long baseline campaign. We found that
these CPDs are detectable in all 3 signposts with ALMA Cycle 3
capabilities for both, 1 and 5~$M_{\rm Jup}$ protoplanets, when
embedded in a locally isothermal disk. On the other hand, in the
pessimistic case of an adiabatic disk the CPD formation was hampered,
hence detectability is scant. Previous radiation hydrodynamic
calculations, show that CPDs are better described by isothermal disks
and that the adiabatic simulation is in fact a rather pesimistic case
\citep[see][their fig. 12]{Ayl2009}.

The immediate vicinity of the planet's Hill sphere, including the CPD
itself, offers an environment for gas-phase physics which produces
distinctive kinematic observational features. Future ALMA long
baseline observations of gas tracers could detect these signposts of
planet formation, and provide not only confirmation of forming planets
but also valuable kinematic information on CPD physics.

\acknowledgments

We thank the referee for a careful review. SP acknowledges financial
support by FONDECYT grant 3140601. Financial support was provided by
Millennium Nucleus RC130007 (Chilean Ministry of Economy), and
additionally by FONDECYT grants 1130949, 1141175, 3140634. AD and PR
acknowledge ALMA-CONICYT grants 31120007 and 31120006. SM acknowledges
CONICYT-PCHA 2014-22140628. MM acknowledges CONICYT-Gemini grant
32130007. The Geryon2 cluster housed at Centro de Astro-Ingenieria UC
was used for the SPH calculations. The BASAL PFB-06 CATA, Anillo
ACT-86, FONDEQUIP AIC-57, and QUIMAL-130008 provided funding for
improvements to the Geryon2 cluster.

\end{document}